# The Mass, Fake News, and Cognition Security

Bin Guo*[1], Yasan Ding[1], Yueheng Sun[2], Shuai Ma[3], Ke Li[1]

1. Northwestern Polytechnical University, China

2. Tianjin University, China

3. Beihang University, China

guob@nwpu.edu.cn

*Abstract*—The wide spread of fake news in social networks is posing threats to social stability, economic development and political democracy etc. Numerous studies have explored the effective detection approaches of online fake news, while few works study the intrinsic propagation and cognition mechanisms of fake news. Since the development of cognitive science paves a promising way for the prevention of fake news, we present a new research area called Cognition Security (CogSec), which studies the potential impacts of fake news to human cognition, ranging from misperception, untrusted knowledge acquisition, targeted opinion/attitude formation, to biased decision making, and investigates the effective ways for fake news debunking. CogSec is a multidisciplinary research field that leverages knowledge from social science, psychology, cognition science, neuroscience, AI and computer science. We first propose related definitions to characterize CogSec and review the literature history. We further investigate the key research challenges and techniques of CogSec, including human-content cognition mechanism, social influence and opinion diffusion, fake news detection and malicious bot detection. Finally, we summarize the open issues and future research directions, such as early detection of fake news, explainable fake news debunking, social contagion and diffusion models of fake news, and so on.

.

## I. Introduction

The rapid popularization and development of social networks have created a direct path from content producers to consumers, changing the way users access information, debate, and form their opinions. Instead of accessing news from traditional and curated mechanisms, such as news broadcast or daily news programs, people are turning to social media platforms which expose them to a broader range of opinions and information about the issues of the day. The growth of *social media* has changed patterns of consumption and exposure to a variety of news deliberately and incidentally, and social media platforms have become a major source of news[1], such as Facebook[2], Twitter[3], YouTube[4], Instagram[5] and Snapchat[6]. Although social networks have accelerated the dissemination of information and promoted the communication of people, contemporary social media platforms offer a hotbed of spreading fake news due to their low cost, easy access and high anonymity. A survey conducted by the Pew Research Center shows that nearly 23% of interviewed Americans have ever reposted and shared fake news on social networks[7]. In addition, the existence of social bots, botnets and trolls have also been a severe problem in social media platforms. It is reported that as many as 60 million trolls could be spreading fake news on Facebook [1]. Furthermore, the prevalence of fake news in social networks confuses the audience, creates panic, and seriously affects public safety and mass cognition security [2].

The spread of fake news is posing threats to diverse domains, such as vaccine safety, climate change, political elections, and stock stability [3]. For example, during the U.S. presidential election in 2016, *PolitiFact*, an independent fact checker of political statements, judged 70% of all statements about Donald Trump to be false or mostly false[8] and Trump's supporters were far more likely to consume fake news than Clinton's supporters [4]. Consequently, 'fake news' was named the "word of the year" by Collins Dictionary in 2017 since it has aroused spread concern of the world. In addition to political interference, fake news can also do great damage to social stability. For example, the fake news on social media about Turkish government's implementation of capital controls led to a 20% drop in the lira against the US dollar[9], causing huge economic loss in Turkey. The fake news which claimed that the border between Greece and North Macedonia was open made hundreds of migrants and refugees pour across the Greek border[10]. It further results in the clash between Greek police and migrants. Thus，it can be seen that fake news is one of the current greatest threats to democracy, economy and journalism [5].

In 2018, the *Science* magazine launched a special issue about 'Fake News', where they discussed the conception, network propagation mechanism and social influence of fake news [2, 6]. In [7], Ruths divides the dissemination process of fake news into five key components, consisting of publishers, authors, articles, audience and rumors. Qiu *et al.* [8] find that both information overload and limited attention contribute to the degradation of human's ability to judge news whether fake or true. Lazer *et al.* [2] identify two categories of fake news interventions, including empowering individuals to evaluate the fake news and utilizing platform-based detection and algorithms.

---

[1] https://www.oberlo.com/blog/social-media-marketing-statistics
[2] https://www.facebook.com/
[3] https://twitter.com/
[4] https://www.youtube.com/
[5] https://www.instagram.com/
[6] https://www.snapchat.com/

[7] https://www.journalism.org/2016/12/15/many-americans-believe-fake-news-is-sowing-confusion/
[8] https://www.politifact.com/personalities/donald-trump/
[9] https://www.theguardian.com/world/2018/aug/13/turkey-financial-crisis-l-ira-plunges-again-amid-contagion-fears
[10] https://www.dw.com/en/greek-police-clash-with-migrants-near-north-ma-cedonia-border/a-48240710

An urgent concern is that the development of Artificial Intelligence (AI) technology puts forward higher requirements for fake news identification. The research of fake news will extend from text to high-quality, machine-generated and manipulated images, videos and audios on a massive scale [9]. For instance, Deepfakes [10, 11], creates audios or videos of real people they never said or did by neural networks, which has been widely used to forge politicians' speeches and illegal evidence [12], resulting in hurting public feelings and affecting the political situation seriously.

To summarize, fake news can influence the emotions, opinions, and other cognition activities through human-content interactions. With the idea that some information succeeds due to their content taps into general cognitive preferences [13], it is significant to understand the cognition and dissemination mechanism of fake news before checking the fact. This paper presents a promising research area called "Cognition Security (CogSec)", which aims to *understand the interaction patterns, cognition behaviors, and social influence & diffusion mechanism between human and fake news, and investigates the successful and efficient ways to debunk fake news and maintain human cognition security.*

CogSec is a multidisciplinary field of research that leverages knowledge from social science, psychology, cognition science, neuroscience, AI, and computer science.

In particular, the main contribution of this work are three folds.

- Characterizing the Cognition Security (CogSec) research area, ranging from its concept model and research scope.
- Investigating the main research challenges of CogSec and presenting the state-of-the-art techniques to address these issues.
- Discussing the open issues and future research directions of CogSec.

## II. CHARACTERIZING COGNITION SECURITY

In addition to fake news, there are other types of information spreading on social media platforms that threaten the CogSec, such as *rumor*, *hoax*, *click-bait*, *disinformation*, and *misinformation*. The widely-recognized definitions are summarized in Table 1.

For characterizing the research area of cognition security, this section firstly presents the problem statement about CogSec. In this paper, we follow the definition of *fake news* used in recent papers [18, 19].

DEFINITION 2.1. **Fake news**: *A news article that is intentionally and verifiable false.*

The abundant users of social media platforms generate a massive number of contents based on social interactions. Human interact with such online contents and their perceptions, behaviors, and knowledge are implicitly influenced [20, 21]. We define the human-content interaction as follows.

DEFINITION 2.2. **Human-content interaction**: *Publish, share, like, and comment of online contents (e.g., news, posts, photos, videos, etc).*

We further give the definitions of cognition security and cognition security protection.

DEFINITION 2.3. **Cognition security**: *CogSec refers to the potential impacts of fake news to human cognition, ranging from misperception, untrusted knowledge acquisition, targeted opinion/attitude formation, to biased decision making.*

DEFINITION 2.4. **Cognition security protection**: *CogSec protection is committed to effective intervention to ensure humans' CogSec, including the techniques of cognition mechanism investigation, diffusion pattern mining, early fake news detection, malicious bot detection, and so on.*

Regarding the scale of human-beings the cognition security can affect, it can be categorized into the *individual* level, the *crowd* level, and the *society* level.

Traditional vision of network security [22] mainly emphasizes data and information security, while CogSec focuses on the complex interaction mechanism between human cognition and multimodal content of social media, expanding from the traditional "*machine*" security to "*human-machine*" fusion security, as presented in Table 2.

TABLE I. DEFINITIONS OF SOME TYPES OF MALICIOUS INFORMATION

| Term | Definition |
|---|---|
| *Rumor* | An item of circulating information whose veracity status is yet to be verified at the time of posting. [14] |
| *Hoax* | A deliberately fabricated falsehood made to masquerade as truth. [15] |
| *Click-bait* | A piece of low-quality journalism which is intended to attract traffic and monetize via advertising revenue. [16] |
| *Disinformation* | Fake or inaccurate information which is intentionally false and deliberately spread. [17] |
| *Misinformation* | Fake or inaccurate information which is unintentionally spread. [17] |
| *Fake news* | A news article that is intentionally and verifiable false. [18] |

TABLE II. DIFFERENCES WITH OTHER CONCEPTS

| Term | Research Focus | Security Paradigm |
|---|---|---|
| Network security | Data and content security | Machine security |
| Cognition security | The interaction and cognitive mechanism between human and contents in the cyberspace | Human-machine security |

Recently, there have been several related studies and important findings regarding this research field, representative ones as presented below.

*(1) Echo chambers* [23-25]. It traps users by only exposing them to opinions and beliefs they are already in agreement with [26]. Echo chambers is compounded by the rise of algorithmic news recommendation and content filtering [27], which makes users always browse their favorite information and implicitly influences users' cognitive behaviors. For example, Barberá *et al.* [28] observe that information is mainly exchanged among users with similar ideological preferences in the case of political issues. Similarly, Quattrociocchi *et al.* [29] demonstrate that such echo chambers really reinforce selective exposure and group polarization. People tend to only concentrate on confirming claims and ignore obvious objections, because they focus on their preferred information. Moreover, Zajonc *et al.* [30] assume that the perceived accuracy of false information increases linearly with the frequency of exposure to the same false information, which means that fake news repeatedly appearing in echo chambers may gradually be accepted as true news. Above all, highly homogeneous echo chambers in social networks can decrease people's ability to identify fake news and increase their misperceptions, contributing to spreading false information [31].

*(2) Online gatekeepers* [32, 33]. It refers to information controller (information selection, deletion, manipulation or integration etc.) in the process of information dissemination [34]. Xu *et al.* [35] observe that users in social networks are highly likely to become gatekeepers. In [36], Garimella *et al.* explore the role of gatekeepers in the creation of echo chambers in case of political news, and they find these gatekeepers usually have lower clustering coefficient. Although online gatekeepers consume information with different viewpoints, they tend to share only a certain viewpoint to strengthen the homogeneity of target community and form a closed field of public opinion, which contributes to the dissemination of fake news [37]. Therefore, effective use of gatekeepers to prevent the spread of fake news needs to be further studied.

*(3) Media bias* [38, 39]. It is one type of cognitive bias, which means that journalists are unable to report news events fairly and objectively due to their partial opinions [40]. As Jamieson *et al.* [41] recognize, the news media does not just report the facts, but is often affected by government influence, targeting at audiences' preference, sponsor pressure and so on. Under the comprehensive impact of various aspects as well as the purpose of chasing headlines, media outlets often release claims without thorough verification, which provides an opportunity for the spread of fake news. Puglisi [42] finds that the *New York Times* may lean democratic. Besides, Gerber *et al.* [43] estimate that voters who read the *Washington Post* regularly are 8% more likely to vote democratic candidate in the 2005 governor election in Virginia. Many media researchers fear that unregulated media will have a major impact on our society [44], but competition among different media outlets can eliminate ideological bias in some cases [45].

*(4) The spread of fake news* [46, 47]. There are many factors that contribute to the spread of fake news, such as cognitive limitation of readers [48], usability of social media platforms [49], and demographics of audiences [50]. Some studies have been carried out on the propagation characteristics and structures of fake news. For example, DiFonzo *et al.* [51] find that rumors containing negative emotions are more likely to be spread. Guess *et al.* [52] state that conservatives are more likely to share fake news and that Facebook accounts over 65 years old spread about seven times as much fake news as the young during the 2016 US presidential election. Budak *et al.* [53] demonstrate that the popularity of fake news is the result of news production and consumption. They further find that male voters are more impressed by fake news publishers.

III. KEY RESEARCH CHALLENGES AND TECHNIQUES

Having characterized the concepts of CogSec and reviewed some related studies, this section investigates some key research challenges and techniques of this research area, including *human-content cognition mechanism*, *social influence and opinion diffusion*, *fake news detection*, and *malicious bot detection*.

*A. Human-Content Cognition Mechanism*

Understanding the mechanism that people share, repost, and agree of online contents is critical to protect their cognitive security. A thorough understanding of the mechanisms should rely on knowledge from psychology, cognition science, and neuroscience [54].

*(1) Personality, content sharing, and debunking.* Interpersonal social interaction, centered on content sharing, enables information to spread efficiently [55]. Actually, content sharing behaviors among users in social networks, such as *publish*, *repost*, and *like*, will gradually affect the reach and influence of news [56]. There are several studies that aim to learn information sharing mechanism in social media. For instance, Scholz *et al.* [57] present a neurocognitive framework to understand mechanisms under information sharing. Based on the *New York Times* health news articles dataset, they find that the core functions of sharing relate to both self-expression and social bond strengthen. Hodas *et al.* [58] reveal a systematic link between personality type and mood, brain response, and the type of content people choose to share online. They observe that users' preferences might be predicted from both personality and transitory mood state. In [59], Falk *et al.* focus on neural responses of information consumers' brains. They find that individuals are more capable of spreading their opinions to others, thus generating greater mentalizing-system activity in the initial process of information sharing.

Some works predict content reposts in social networks. For example, Hu *et al.* [60] predict the popularity of pictures and their diffusion paths in social networks based on *Diffusion-LSTM*, a memory-based deep recurrent neural network model. A combination of user social features and image features is used to characterize individual reposting behaviors. Similarly, Zhang *et al.* [61] propose an attention-based deep neural network to combine contextual and social context information for *retweet* behavior prediction.

In [62], Lewandowsky *et al.* observe audiences' memories for misinformation and study the role of cognitive factors in misinformation debunking. They further divide human cognitive problems in the face of misinformation into four categories, including *continued influence effect*, *familiarity backfire effect*, *overkill backfire effect*, and *worldview backfire effect*, which provides the theoretical basis and suggestions for CogSec protection.

*(2) Neuroscience in human-content interaction*. Neuroscience has also been widely used in many research areas (e.g., healthcare [63], intelligent control [64, 65], artificial intelligence [66], economics [67] etc.) related to human-computer interaction. As presented by Poldrack *et al.* [68], the use of new tools, e.g., Electroencephalography (EEG), functional Magnetic Resonance Imaging (fMRI), and Magnetoencephalography (MEG), for imaging and manipulating the brain will continue to advance our understanding of how the human brain gives rise to thought and action.

Regarding CogSec, neuroscience has been previously used for understanding human-content interaction. Some efforts have been conducted to understand/predict population-level behaviors/preferences (e.g., ratings and sharing in social media) based on small groups of individuals' neural responses. For example, researchers test the possibility of using fMRI to predict the relative popularity of music[11]. Dmochowski *et al.* [69] find that naturalistic stimuli (viewing multimedia contents) evoke highly reliable brain activities across viewers. Falk *et al.* [70] further conclude that neural responses of a small group of individuals can be used to predict the behavior of large-scale populations. In particular, neural activities in a medial prefrontal region of interest which are previously associated with individual behavior change can predict the population response. Hasson *et al.* [71] report the unexpected finding that brains of different individuals show a highly significant tendency to act in unison during free viewing of a complex scene such as a movie sequence. In [72], Adolphs identifies a series of neural structures involved in users' perceptions and judgements of content stimuli, and analyzes humans' ways of reasoning and decision-making. In general, neuroscience provides the theoretical basis for understanding human-content interaction, and has practical significance for the protection of public CogSec.

*B. Social Influence and Opinion Diffusion*

The study of social influence and opinion diffusion in social networks has a long tradition in the social, physical, and computational sciences. For example, there have been numerous studies on opinion formation [73, 74] and influence maximization models [75]. Here, we review the related studies about the spread of fake news.

*(1) Social influence and contagion*. The concept of social contagion has expanded from the initial epidemic transmission to the process of information dissemination across social networks, such as political views [76], emotional changes [77], fashion trends [78], and financial decisions [79]. Some works measure the influence of opinions in social networks, aiming to make information far-reaching. For example, Morone *et al.* [80] introduce percolation theory [81] to social network influential node discovery and find that a large number of weakly-connected (low-degree) nodes can be optimal influencers. Amati *et al.* [82] utilize *degree*, *closeness*, *betweenness* and *PageRank*-centrality of nodes in Dynamic Retweet Graph [83] to find the most influential users in Twitter. In [84], Qiu *et al.* propose *DeepInf*, a deep learning-based influence prediction framework, which learns users' latent social representation to evaluate their social influence by incorporating *network embedding*, *graph convolution*, and *attention* mechanism.

Some studies concentrate on the contagion and persuasion mechanisms of messages in social networks. For instance, Ugander *et al.* [85] find that whether social network users will be infected depends on the number and structure of their interrelated components, rather than the actual size of the community. Therefore, different social environments and influences represented by target users' neighbors can be considered as the driving mechanism of social contagion. In [86], Kramer *et al.* prove that each user's emotions can be affected by other users in Facebook, which provides an experimental basis for massive-scale social influence and contagion. Abebe *et al.* [87] study the process of information contagion from the perspective of changes in people's psychological sensitivity to persuasion. They further propose a dynamic model of social opinions that comprehensively utilizes the maximization and minimization of crowd opinions for influencing social opinions.

*(2) Spreading models/mechanisms*. As Ratkiewicz *et al.* [88] state, the early stages of the diffusion of rumors tend to show pathological patterns. Thus, some work has studied the spreading mechanisms and modes of online information to provide guidance for CogSec protection. For example, Friggeri *et al.* [89] track the propagation of thousands of rumors appearing on Facebook. They find that rumor cascades run deeper in the social network than normal sharing cascades. Vosoughi *et al.* [6] report that fake news is more novel than real news, suggesting that people are more willing to spread novel information. Besides, the true information usually evokes the users' sadness, happiness, and trust, while fake news often triggers public surprise, fear, and disgust. Similarly, Peng *et al.* [90] find that users are more delight to hear positive gossip and more annoyed to hear negative gossip of themselves, compared with celebrities and their friends. Vicario *et al.* [31] find that misinformation in social networks often leads to homogeneous and polarized communities and propose a data-driven percolation model of misinformation spreading, which demonstrates that homogeneity and polarization are the determinants of predicting the size of information cascade.

Several works have been carried out on the opinion dynamics based on influence mechanism in social networks, which can be divided into discrete models [91, 92] and continuous models [93]. For instance, aiming at understanding the vulnerability of social networks and increasing users' resilience to fake news, Wang *et al.* [94] propose a multivariable jump diffusion guidance framework, which models the dynamics of opinions and guides public opinions to the desired state. Martins *et al.* [95] propose an opinion diffusion model, *CODA*, in which different opinions of users are regarded as discrete variables and each opinion is modeled as continuous opinion function. Target users decide whether to change their own opinions or not based on Bayesian descriptions of their neighbor opinions. In [96], Yang *et al.* design a role-aware information diffusion model (RAIN), which characterizes the interaction between users' social roles and their influence on the spreading of information.

---

[11] https://www.sciencemag.org/news/2011/06/can-brain-scans-predict-music-sales

*C. Fake News Detection*

Since fake news has a great impact on social stability, economic development, and political democracy, it is imperative to study efficiently automatic fake news detection technology [19]. Recently, there have been several efforts on fake news detection, which can be divided into *content-based*, *social context-based*, and *deep learning-based* methods.

*(1) Content-based methods*, which often rely on unique writing styles or language features in news content (e.g., lexical features, syntactic features, and topic features) [97, 98]. For example, Castillo *et al.* [99] calculate a series of linguistic features to evaluate Twitters' credibility, including the average number of words, URL links, the number of positive words etc. Potthast *et al.* [100] propose a meta-learning model to detect fake news, which utilizes differences in writing styles between the truth and fake news. Hu *et al.* [101] propose a spammer detection method based on sentiment information.

*(2) Social context-based methods*, which mainly focus on the characteristics of human-content interactions, such as *user profiling*, *reposts*, *comments*, *stances*, and *likes* etc. For example, Tacchini *et al.* [102] estimate that social media platform posts can be detected as hoax utilizing netizens' *like* behaviors. Ma *et al.* [103] make use of the temporal patterns of social context features to detect online rumors. In [104], Jin *et al.* propose a credibility propagation network model for rumor detection by mining supporting or opposing opinions in microblogs. Yang et al. [105] propose an unsupervised fake news detection model, incorporating the authenticity of news, users' reputation, and users' viewpoints on target news event.

*(3) Deep learning-based methods*, which aim to learn latent representations of fake and real news accurately for further detection. Existing deep learning-based detection methods mainly apply convolution neural network (CNN) [106] and recurrent neural network (RNN) [107] models. For example, Li *et al.* [108] utilize the Bidirectional GRU model to detect online rumors, based on the observation that both the forward and backward sequences of social posts contain abundant interactive information. Liu *et al.* [109] find that there are obvious differences between the propagation patterns of true news and fake news, and they combine GRU (extracting global features) and CNN (extracting local features) to detect fake news. Ruchansky *et al.* [110] propose the RNN-based fake news detection model, incorporating textual features of news, user response, and the source users. Similarly, Shu *et al.* [111] further explore the social relations among publishers, news and online users [112, 113], and then propose a tri-relationship embedding network, *TriFN*, which models the human-content interactions for fake news detection.

*D. Malicious Bot Detection*

The popularity and openness of social network promote the emergence of social bots with certain autonomous decision-making ability [114]. Like legitimate users, social bots can *make friends*, *post tweets*, *thumb up*, *chat* and so on through program control. Salge *et al.* [115] point out that about 8.5% of Twitter accounts are social bots, engaged in news, events, business communication and other tasks. Most social bots provide convenience for users to exchange information by automatically providing benign news and information, but there are also malicious social bots that can spread rumors and harmful information [114, 116, 117]. Recently, a large number of malicious bot detection methods have been proposed, which can be categorized as *behavior-based*, *content-based*, and *influence-based* methods.

*(1) Behavior-based detection methods.* It is of great value to analyze and mine the behavior data of social bots in existing social networks [118]. Boshmaf *et al.* [119] analyze the differences between social bots and human users in terms of the number of friends, post time interval, post content and account attribute differences, and propose a random forest based social bot detection method. Haustein *et al.* [120] analyze the differences between real Twitter users and social bots in retweeting scientific articles, and find that social bots tend not to be selective in retweeting (involving topics, sources, etc.). In [121], Gilani *et al.* conduct a comparative study on the behaviors of human and social bots in posting and retweeting on Twitter, and find that social bots play a very important role in information transmission, despite their weak overall influence. Besides, Varol *et al.* [122] find that compared with human users, the interaction selection of social bots is more arbitrary and that there are fewer bidirectional connections between them and human users.

*(2) Content-based detection methods,* which focus on determining whether a message posted by a user is a malicious message. Generally, whether the URL in the message content points to the malicious page can be used to determine whether the account that published the message is malicious social bot. For instance, Thomas *et al.* [123] propose a real-time URL detection scheme, which extracts features of related URL pages by visiting each published URLs. What's more, social bots can be detected through changes in the message content features. For example, Egele *et al.* [124] extract 7 content features, model the messages, and then judge whether the messages published later deviate from the created model to detect social bots. In [125], Kudugunta *et al.* propose a LSTM-based bot detection method, incorporating contextual features and accounts' metadata for improving bot detection accuracy. Gao *et al.* [126] find that 63% of the text content of spam messages in Twitter is generated based on templates, and they propose the social bot detection framework, *Tangram*, which divides malicious posts into fields, generates matching templates, and detects more malicious social bots.

*(3) Influence-based detection methods,* which detect social bots on the perspective of social influence. For example, Messias *et al.* [127] conduct comparative studies on analyzing the influence of social bots, and propose their malicious behavior strategies, including regular posting tweets on a certain hot topic, different posting intervals, and attribute integrity. Similarly, Abokhodair et al. [128] analyze the posting behavior, social structure, group behavior characteristics and influence growth process of social bot network. Finally, they find that more human-like behaviors can improve social bot influence. Freitas et al. [129] create 120 different attributes (sex, occupation, etc.) and behavior strategies (active, posting action and interaction) of the social bots for characterizing their infiltration process, and they find that about 20% of the social bots gain more than 100 followers by means of high active interaction and posting behavior.

IV. OPEN ISSUES AND FUTURE RESEARCH DIRECTIONS

Though there have been initial efforts in the field of CogSec, there are still numerous research challenges to be tackled in the future, some of which are discussed below.

*(1) The human cognition mechanism of fake news.* Regarding to CogSec protection, the first thing is to understand the human cognition mechanism of fake news. Acerbi [13] estimates that fake news can be successfully disseminated because it meets the general cognitive preferences of the public, which provides theoretical guidance for preventing the spread of fake news. With the rapid development of neuroscience, several studies have investigated the cognitive patterns of the human brain [57, 130, 131]. For example, Lewandowsky et al. [132] raise the problem of "technocognition" and summarize the ways in which fake news affect the society negatively. In [133], Arapakis et al. propose a measurement model for evaluating the interest changes of users in reading news, which is based on EEG registration of people's neural activity. All in all, the research on the cognition mechanism of fake news correlates to multiple disciplines such as psychology, neuroscience and cognitive science. We still need to explore specific cognitive problems, partially summarized as follows.

- The influence of individual's social cognition on large-scale social behaviors.
- The common features of fake news satisfying users' cognitive preferences.
- The effect of fake news content stimulations (text, pictures, audios or videos) on specific parts of the human brains.
- The impact of social interactions on individual's cognition.
- The change of users' cognitive characteristics with the dissemination of fake news.

*(2) The social contagion and diffusion models of fake news.* Social contagion is a common phenomenon in human society [134], which contributes to opinion dynamics, behavior shaping, and cognitive preferences in social networks. Some works pay attention to modeling the contagion and propagation of information in social networks. For instance, Chang et al. [135] explore how social media marketing persuades users to share information with the purpose of achieving mass cohesion and information diffusion. Huang et al. [136] propose a social contagion model based on introducing a persuasion mechanism into the threshold model. They then estimate that persuasion mechanism improves the influence of information cascade in social networks, and that the effect of persuasion is often more significant in heterogenous social networks than in homogeneous networks. In the future, there are still numerous issues to be further studied:

- The study of novel information dissemination theories which introduce the users' cognition preferences, timeliness of information, and social roles of individuals, etc.
- The evaluation of influential users on social networks for maximizing the impact of information dissemination.
- The fast influence maximization mechanisms of true information to the mass after fake news debunking.

*(3) Early detection of fake news.* Information on social networks usually has a short life span, averaging less than three days, and fake news always spread like viruses with a few minutes [89, 137]. Actually, detection methods based on aggregation features (e.g., propagation characteristics, etc.) are difficult to achieve better performance on early detection [138]. Therefore, the early detection of fake news is an important issue. Some works attempt to identify fake news at their early spreading stage. For example, Zhao et al. [139] find that queries and objections in users' comments contribute to early detection of rumors. Chen et al. [140] find that users tend to comment differently in different rumors' spreading process and propose an RNN-based rumor detection model with attention mechanism for early detection. In [141], Sampson et al. utilize implicit linkages for acquiring additional information from several related events to deal with the problem that less data is available in the early detection of fake news. Although several studies have been conducted on the early detection of fake news, the performances of them still need to be improved.

*(4) Explainable fake news debunking.* Existing automatic fake news detection models [99, 107] usually just give the testing results, with little decision-making basic explanations. However, the explanation in fake news debunking or the transparency of detection models is essential, which contributes to users' trust in detection results, fusion of human-machine intelligence, and further prevention of the spread of fake news. Some studies utilize the attention mechanism [140, 142] and graph models [143, 144] for explainable fake news debunking. For instance, Popat et al. [145] propose an automatic end-to-end fake news detection model combined with external evidence articles, *DeClarE*, based on Bidirectional LSTM with attention. Similarly, Guo et al. [146] introduce social contexts into rumor detection via attention mechanism to enhance the interpretability of detection models, based on hierarchical LSTM. Gad-Elrab et al. [147] propose a framework for generating explanations of candidate facts, incorporating knowledge graphs and texts, which provides reference for fake news detection. In general, explainable fake news debunking needs to explore more practical models with the development of interpretable machine learning (IML) [148, 149], such as probabilistic graphical model (PGM) [150], knowledge graph based on complex rules [151], and other mechanisms.

## V. CONCLUSION

In the context of the spread of fake news in social networks, we present a novel research issue, named Cognitive Security (CogSec). In order to characterize the CogSec, we propose some relevant definitions and review several related findings, including echo chambers, online gatekeepers, media bias etc. We further investigate the key research challenges and techniques of CogSec, which can be categorized into human-content cognition mechanism, social influence and opinion diffusion, fake news detection, and malicious bot detection. The study of CogSec is still at its early stage, and there are still numerous challenges and open issues to be addressed by AI researchers, social and neuroscience scientists, as well as security engineers.

## ACKNOWLEDGEMENT

This work was partially supported by the National Natural Science Foundation of China (No. 61772428,61725205).